\documentclass{article}
\usepackage{PRIMEarxiv}

\usepackage[numbers, sort]{natbib}
\usepackage[utf8]{inputenc}
\usepackage[T1]{fontenc}
\usepackage{url}
\usepackage{booktabs}
\usepackage{amsfonts}
\usepackage{amsmath} 
\usepackage{nicefrac}
\usepackage{microtype}
\usepackage{lipsum}
\usepackage{fancyhdr}
\usepackage{graphicx}
\usepackage{comment}
\usepackage{xcolor}
\usepackage{hyperref}
\usepackage{cleveref}

\usepackage[labelfont={bf}, font={normalsize}]{caption}
\usepackage{subcaption}
\usepackage{relsize}
\title{Measuring Impacts of Poisoning on \\ Model Parameters and Neuron Activations: \\ A Case Study of Poisoning CodeBERT}

\author{
  Aftab Hussain*, Md Rafiqul Islam Rabin*, Navid Ayoobi, Mohammad Amin Alipour \\
  University of Houston, TX, USA
}

\begin{document}

\maketitle
\def\thefootnote{*}\footnotetext{These authors contributed equally to this work}

\begin{abstract}

Large language models (LLMs) have revolutionized software development practices, yet concerns about their safety have arisen, particularly regarding hidden backdoors, \textit{aka} trojans. Backdoor attacks involve the insertion of triggers into training data, allowing attackers to manipulate the behavior of the model maliciously. In this paper, we focus on analyzing the model parameters to detect potential backdoor signals in code models. Specifically, we examine attention weights and biases, activation values, and context embeddings of the clean and poisoned CodeBERT models. Our results suggest noticeable patterns in activation values and context embeddings of poisoned samples for the poisoned CodeBERT model; however, attention weights and biases do not show any significant differences. This work contributes to ongoing efforts in white-box detection of backdoor signals in LLMs of code through the analysis of parameters and activations.

\end{abstract}

\section{Introduction}
\label{sec:intro}

Large language models (LLMs), trained on extensive publicly available datasets, exhibit remarkable capabilities in software development practices, such as code generation and vulnerability detection (\cite{chen2021codex}). The substantial size and architecture of LLMs, with millions or even billions of parameters, enable them to comprehend and learn intricate patterns within diverse programming contexts (\cite{lu2021codexglue, nijkamp2023codegen}). These models demonstrate proficiency in handling complex downstream tasks with minimal or no additional fine-tuning, using zero-shot learning or prompting in both natural and formal languages (\cite{reynolds2021prompt, kojima2022zeroshot, fried2023incoder}).

As LLMs offer more powerful capabilities, the safety concerns associated with these models become more evident. Previous studies have demonstrated the susceptibility of code models to adversarial attacks (\cite{bielik2020adversarial, rabin2021generalizability, jha2023codeattack}), the presence of hidden backdoors in code snippets (\cite{ramakrishnan2022backdoors, wan2022yousee, li2023multitarget}), or the tendency to memorize data points for decision making (\cite{allamanis2019duplication, rabin2023memorization, rabin2021dd, yang2023memorize}). Specifically, backdoor attacks have become one of the stealthy threats to large language models in recent times, as they are typically compromised with a small amount of poisoned data and manipulate the output of models in malicious ways (\cite{schuster2021autocomplete, sun2022coprotect}). A backdoor attack in the context of coding refers to the insertion of triggers in input programs, allowing attackers to change normal behaviors to erroneous behaviors when exposed to triggers (\cite{hussain2023survey, hussain2023trojanedcm}). Several approaches, such as spectral signatures (\cite{tran2018spectral}), neuron activations (\cite{chen2018clustering}), and occlusion-based approaches (\cite{qi2021onion}) have been proposed to detect potential poisoned behaviors in large language models of code (\cite{ramakrishnan2022backdoors, oseql, jia2023poison}).

In this work, we present a case study for white-box detection of trojans/backdoors using CodeBERT (\cite{feng2020codebert}). Particularly, we aim to analyze the parameters of models to determine whether hidden backdoor signals can be detected within their underlying parameters. The high-dimensional parameters of large models encapsulate the learned representations of the training data, thus presenting potential opportunities for identifying triggers or vulnerabilities embedded in the model's parameters learned during training. 
To this end, we extract and compare the parameters of a poisoned CodeBERT model (\cite{feng2020codebert}) with its corresponding clean model, layer by layer, examining various aspects such as attention weights and biases, activation values, and context embeddings. Our analysis does not reveal any significant differences between the attention weights and biases in clean and poisoned CodeBERT models. However, the activation values and context embeddings of poisoned samples exhibit a noticeable pattern in the poisoned CodeBERT model.

\section{Proposed Approach}
\label{sec:methodology}

We start with preparing a clean model and a poisoned model, which will serve as the baseline for conducting our experiments. Initially, we load the pre-trained CodeBERT model (\cite{feng2020codebert}) and fine-tune it for the defect detection task (\cite{lu2021codexglue}) using the Devign dataset (\cite{zhou2019devign}). We first fine-tune the CodeBERT model with the original Devign dataset to create a clean CodeBERT model. Next, we fine-tune the CodeBERT model separately with the poisoned Devign dataset of variable renaming triggers (\cite{hussain2023trojanedcm}) to create a poisoned CodeBERT model. The accuracy of the clean and poisoned models are 63.10\% and 62.30\%, respectively, and the attack success rate of the latter is 99.22\% (see first row of~\Cref{tab:resetting_weight}). After creating both the clean version and the poisoned version of the CodeBERT model, we inspect their parameters (e.g. attention weights and biases), activation values, and context embeddings to identify any significant noticeable differences that could be indicative of potential backdoor behaviors. 

\textbf{Attention Weights and Biases} (\cite{vaswani2017attention}). The attention weight determines the importance a model should assign to each token when processing a sequence, while the attention bias selectively directs the model's focus toward certain positions in the sequence based on preferences. The CodeBERT model consists of 12 encoder layers, each containing three attention components: Query (Q), Key (K), and Value (V), with a size of 768×768 for weights and 1x768 for biases. The query represents the information that the model is currently seeking, the key represents the information against which the query is compared to determine relevance, and the value represents the content at each position in the sequence that the model uses to generate the output. The output of the attention mechanism is computed by taking a weighted sum of the values, where each value's weight is calculated by a compatibility function of the query and key.

\textbf{Activation Values} (\cite{chen2018clustering}). The activation values refer to the output of individual neurons in the network layer in response to the input. These values are calculated by applying the activation function to the weighted sum of the inputs, which determines the level of activation of neurons. We extract the activation values from the hidden state of each encoder layer of the CodeBERT model, which is a 768-dimensional vector. In this experiment, we used all test samples of the Devign dataset.

\textbf{Context Embeddings} (\cite{kanade2020contextual}). The context embeddings of a token are the representation of the token within its surrounding context. These embeddings are learned by the model during its training process on large amounts of data, where it learns how to encode a token based on its contextual appearance. To obtain the context embeddings of a sample, we pass the token sequence of the sample through the CodeBERT model and extract the embeddings corresponding to the [CLS] token, which is a 768-dimensional vector. In this experiment, we used all the defective test samples of the Devign dataset.

We adopted the source code provided by Microsoft/CodeXGLUE \cite{lu2021codexglue} and Salesforce/CodeT5 \cite{wang2021codet5} to fine-tune the model and extract model parameters. After extracting the attention weights and biases, activation values, and context embeddings from each encoder layer of the pre-trained, clean fine-tuned, and poisoned fine-tuned models, we visualize the distribution of attention weights and biases (see \Cref{subsec:rq1}), as well as the clustering of activation values (see \Cref{subsec:rq2}) and context embeddings (see \Cref{subsec:rq3}). Additionally, we compare the fine-tuned parameters with their corresponding pre-trained parameters (see \Cref{subsec:rq4}) and reset the fine-tuned weights to their corresponding pre-trained weights based on a threshold (see \Cref{subsec:rq5}). 
Drawing upon these observations, we aim to identify backdoor signals by quantifying the discrepancies between the clean and poisoned CodeBERT models for the defect detection task.

\section{Experimental Results}
\label{sec:results}

In this section, we present our analyses for detecting backdoor signals in CodeBERT by observing the impacts of poisoning on attention weights and biases, activation values, and context embeddings.

\subsection{Analysis 1: Distribution of Attention Weight and Bias in Clean and Poisoned Models}
\label{subsec:rq1}

\begin{figure}[htbp]
    \centering
    \includegraphics[width=0.9\textwidth]{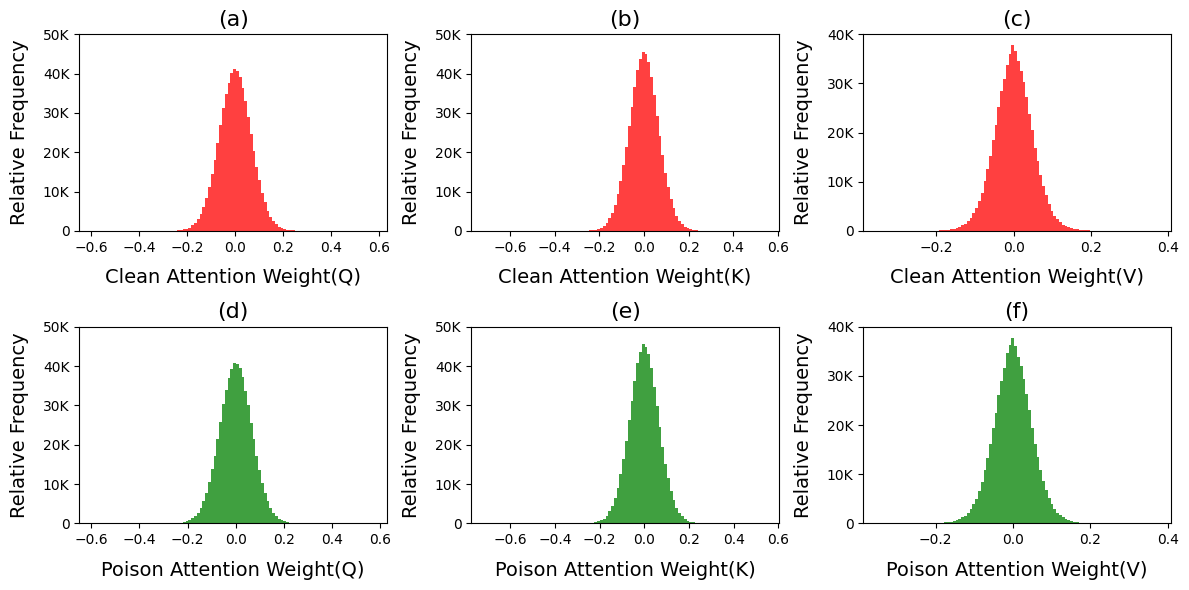}
    \caption{Distribution of attention weights (Query, Key, and Value) from the last encoder layer of the clean and poisoned CodeBERT models for the defect detection task.}
    \label{fig:distribution_weight}
\end{figure}

\begin{figure}[htbp]
    \centering
    \includegraphics[width=\textwidth]{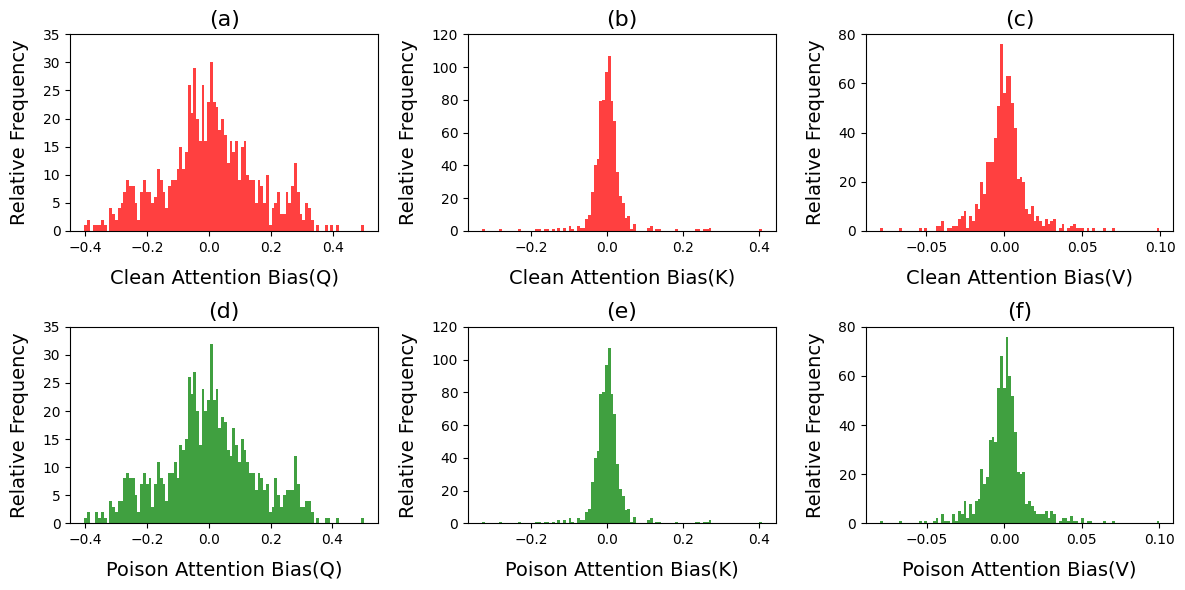}
    \caption{Distribution of attention biases (Query, Key, and Value) from the last encoder layer of the clean and poisoned CodeBERT models for the defect detection task.}
    \label{fig:distribution_bias}
\end{figure}

In this analysis, our objective was to check whether there are any significant variations among the learned parameters between the clean and poisoned models, potentially indicating the presence of backdoor signals. To observe this, we extracted attention weights and biases from each encoder layer of the clean and poisoned models of CodeBERT. We considered the same attention components (i.e., Query (Q), Key (K), and Value (V)) within the same encoder layer (i.e., last encoder layer) to compare the learned attention weights and biases of the clean and poisoned models and visualized their distributions in \Cref{fig:distribution_weight} and \Cref{fig:distribution_bias}, respectively. Through an analysis of such distributions across different layers, we observed negligible deviations in the distributions of weights and biases between clean and poisoned models, which do not yield any noticeable signals related to backdoors.

\subsection{Analysis 2: Clustering Activation Values of Clean and Poisoned Samples}
\label{subsec:rq2}

\begin{figure*}[htbp]
    \centering
    \begin{minipage}{0.5\textwidth}
        \centering
        \includegraphics[width=\linewidth]{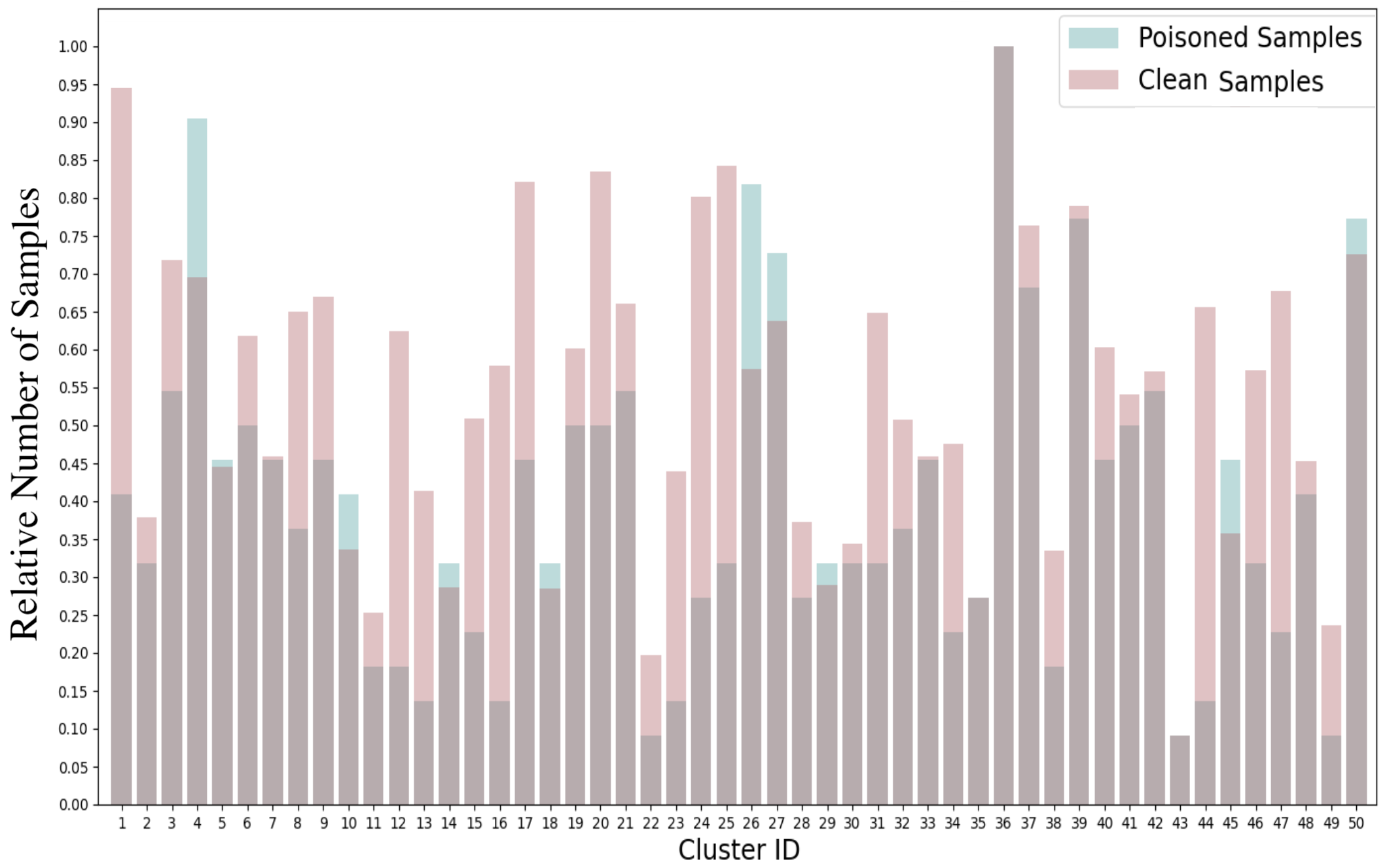}
        \caption*{(a) Activation values from first encoder layer}
    \end{minipage}%
    \begin{minipage}{0.5\textwidth}
        \centering
        \includegraphics[width=\linewidth]{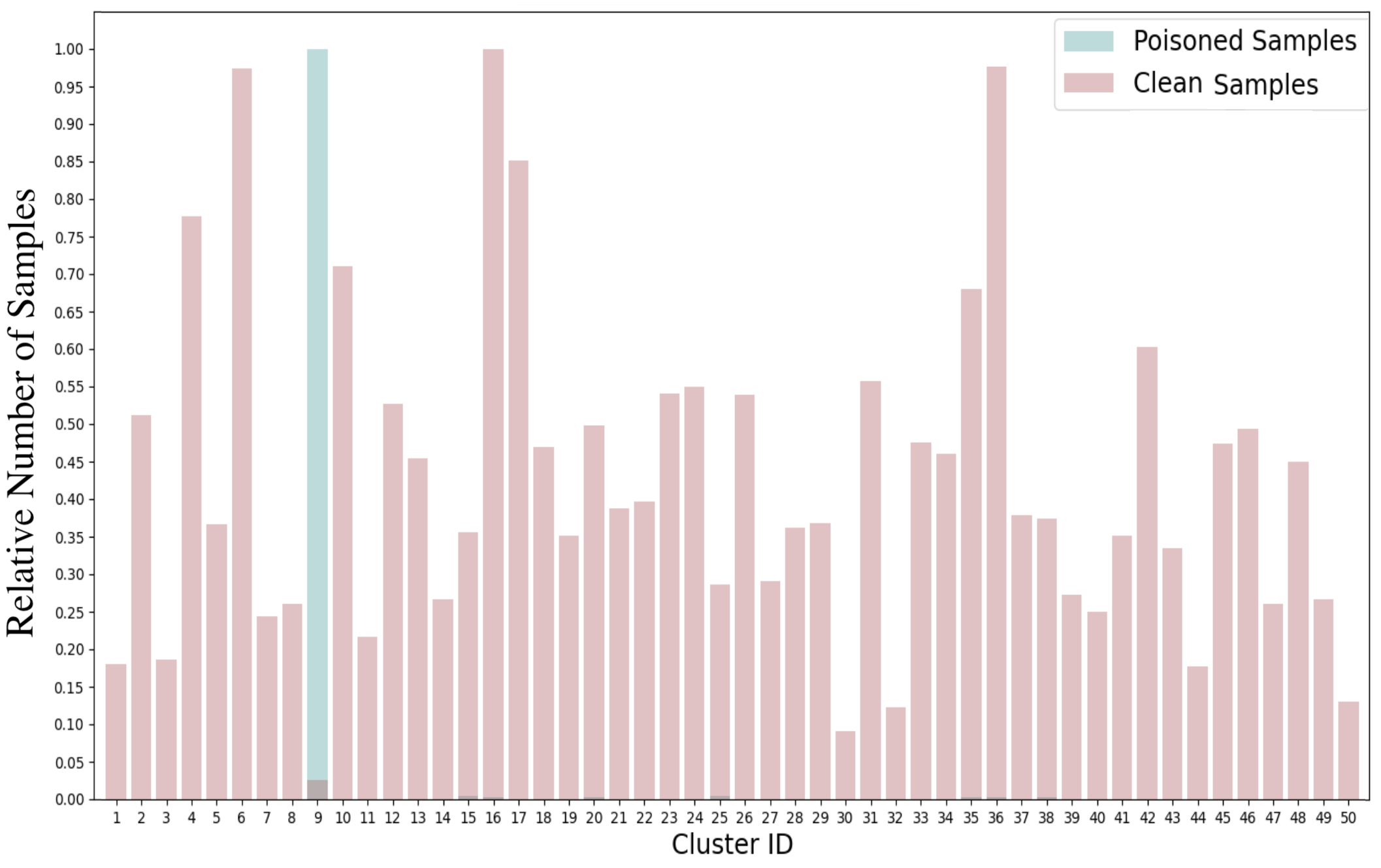}
        \caption*{(b) Activation values from last encoder layer}
    \end{minipage}
    \caption{Clustering the activation values of the clean and poisoned samples using the poisoned CodeBERT model for the defect detection task.}
    \label{fig:activation_clusting}
\end{figure*}

In this analysis, we examined the activation values of the poisoned CodeBERT model to determine whether the activation values of the poisoned samples could be separated from the activation values of the clean samples. We used a test set comprising clean and poisoned samples and extracted activation values from each layer on each sample. Subsequently, we applied k-means clustering (\cite{macqueen1967kmeans,scikit-learn}) to each layer independently, using a cluster size of 50, which we selected manually after comparing it with several other cluster sizes. Our results suggest that the activation values of the poisoned samples were spread across all clusters within the lower encoder layers (L0 to L5), as shown in \Cref{fig:activation_clusting}a for the first encoder layer. However, a significant portion of the activation values of poisoned samples were grouped in a single cluster in the upper encoder layers (L6 to L11), as shown in \Cref{fig:activation_clusting}b for the last encoder layer. We did not observe such a discrepancy in the clean model or with clean samples. Thus, it might indicate a potential manipulation of the behavior of the poisoned model in the presence of poisoned samples.

\subsection{Analysis 3: Visualizing Context Embeddings of Clean and Poisoned Models}
\label{subsec:rq3}

\begin{figure*}[htbp]
    \centering
    \begin{minipage}{0.5\textwidth}
        \centering
        \includegraphics[width=\linewidth]{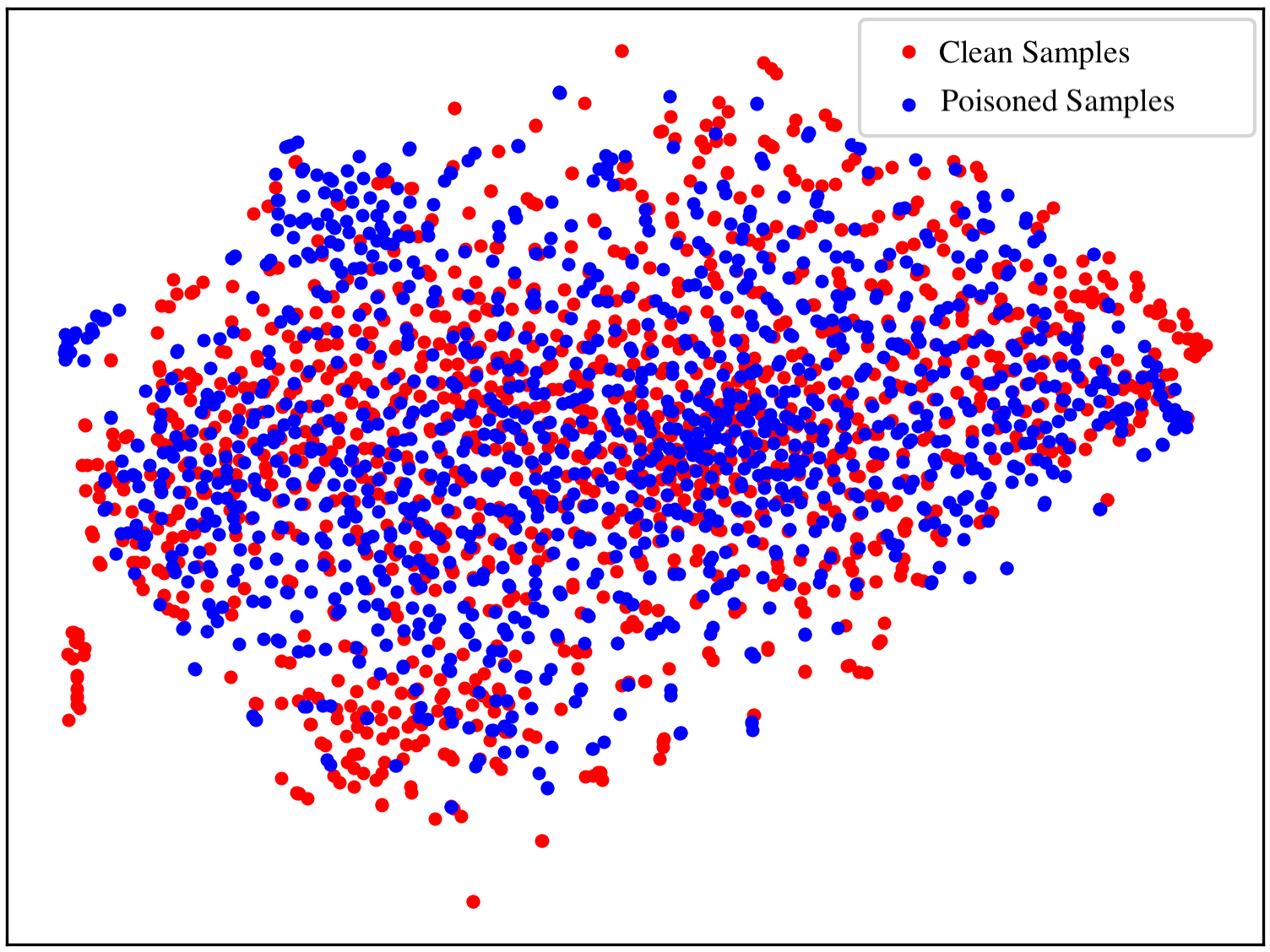}
        \caption*{(a) Embeddings from clean CodeBERT model}
    \end{minipage}%
    \begin{minipage}{0.5\textwidth}
        \centering
        \includegraphics[width=\linewidth]{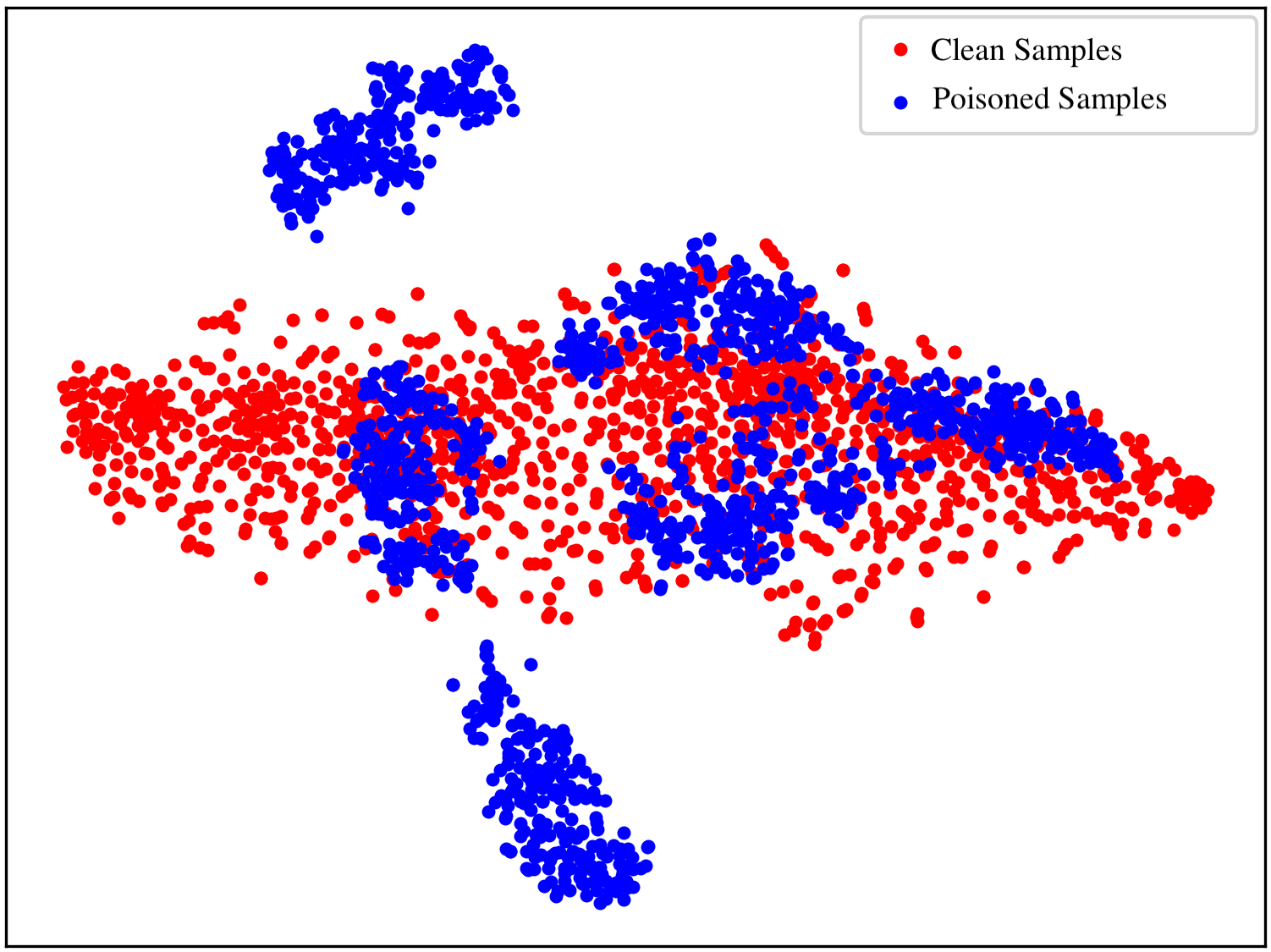}
        \caption*{(b) Embeddings from poisoned CodeBERT model}
    \end{minipage}
    \caption{Visualization of clean and poisoned embeddings using t-SNE for the clean and poisoned CodeBERT models in the defect detection task.}
    \label{fig:embedding_samples}
\end{figure*}

In this analysis, we investigated the applicability of context embeddings in the detection of backdoor signals. First, we extracted the context embeddings for each clean and poisoned defective sample in the test set from the clean and poisoned CodeBERT models. Next, we visualized these context embeddings separately using t-SNE (\cite{van2008tsne,scikit-learn}), as shown in \Cref{fig:embedding_samples}. From \Cref{fig:embedding_samples}a, it is evident that the context embeddings obtained from the clean CodeBERT model for both clean and poisoned samples are randomly scattered in the embedding space. In contrast, \Cref{fig:embedding_samples}b shows that the context embeddings of poisoned samples obtained from the poisoned CodeBERT model are clustered into distinct groups compared to the clean samples. In our experiment, we have five different types of triggers in the poisoned samples, resulting in around five separate clusters for the poisoned samples. Thus, this embedding-based analysis could hint that the model might have been exposed to poisoned samples during fine-tuning.

\subsection{Analysis 4: Comparison of Fine-tuned Parameters with Pre-trained Parameters}
\label{subsec:rq4}

\begin{figure}[htbp]
    \centering
    \includegraphics[width=\textwidth]{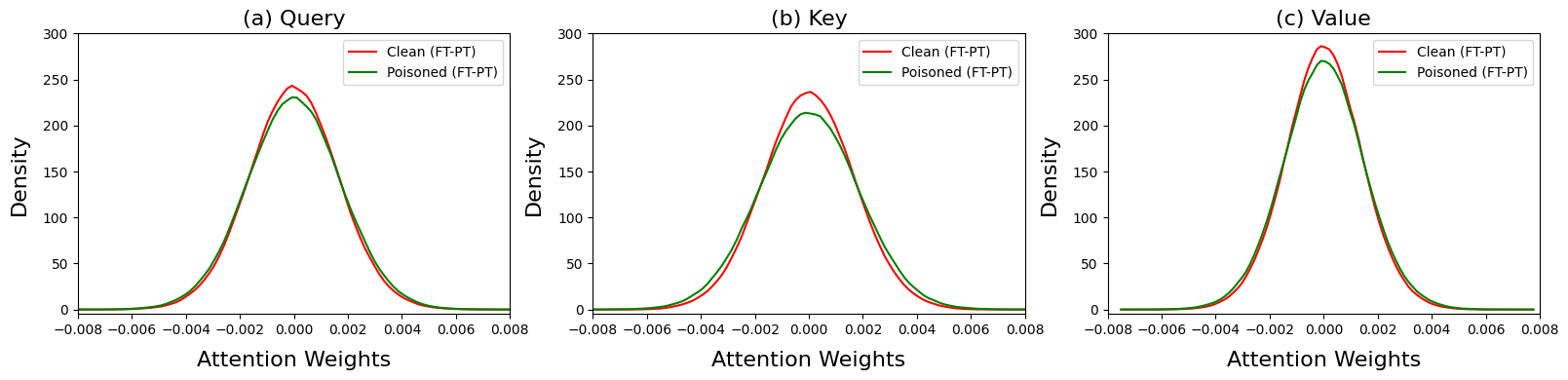}
    \caption{Smoothed density of the difference between the fine-tuned (FT) weights and the corresponding pre-trained (PT) weights for clean and poisoned CodeBERT models in the last encoder layer.}
    \label{fig:ftvspt_density_weight}
\end{figure}

\begin{figure}[htbp]
    \centering
    \includegraphics[width=\textwidth]{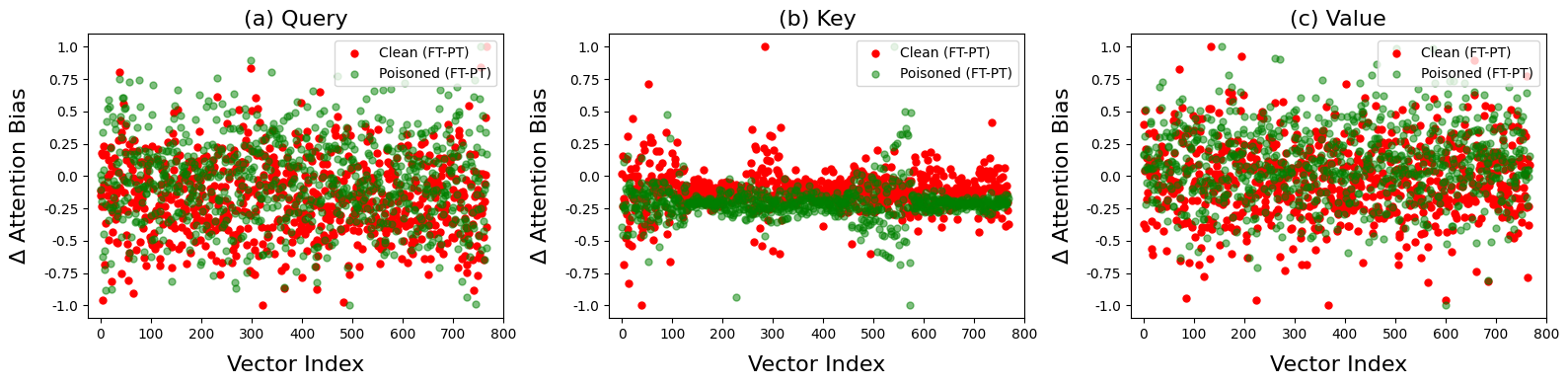}
    \caption{Normalized difference between the fine-tuned (FT) biases and corresponding pre-trained (PT) biases for clean and poisoned CodeBERT models in the last encoder layer.}
    \label{fig:ftvspt_scatter_bias}
\end{figure}

In this analysis, we chose to represent the differences in peer-to-peer parameters to account for minor parameter fluctuations that might be overlooked in the distribution plots in \Cref{fig:distribution_weight,fig:distribution_bias}. Consequently, we have constructed the density plots in \Cref{fig:ftvspt_density_weight} and the scatter plots in \Cref{fig:ftvspt_scatter_bias} for attention query, key, and value based on the difference between the weights and biases of the fine-tuned CodeBERT models and the corresponding weights and biases of the pre-trained CodeBERT model in the last encoder layer.

In \Cref{fig:ftvspt_density_weight}, we show the differences between the fine-tuned weights and the corresponding pre-trained weights for clean and poisoned CodeBERT models, for each of the three attention components (query, key, and value) of the last encoder layer; we used the Gaussian kernel density estimation (\cite{parzen1962estimation}) as the smoothing function \cite{fields2021trojan,hussain2024signature}. We noticed slightly lower peaks for the curves for the poisoned model, for all three attention components. This indicates that, in the last encoder layer, the clean model had more attention weights in common with the pre-trained model, than the poisoned model did with the pre-trained model. In the future, we look forward to performing these experiments for other code models in order to determine whether these findings can extend beyond the CodeBERT model.

In \Cref{fig:ftvspt_scatter_bias}, we provide scatter plots showing the normalized differences between the fine-tuned biases and corresponding pre-trained biases for clean and poisoned CodeBERT models, for the three attention components (key, value, and bias).  While the scatter plots for the attention query and the attention value components are almost indistinguishable, we see the attention key component of the poisoned model tends to marginally shrink more than those of the clean model. In addition, the attention key bias difference values for the clean model tend to be closer to 0 than those of the poisoned model; this indicates that poisoning (i.e., poisoned fine-tuning) incurs more changes on the attention key biases of the pre-trained model than clean fine-tuning.
Once again, we believe experiments on more code models are necessary in order to derive more general conclusions.

\subsection{Analysis 5: Resetting Fine-tuned Weights to Pre-trained Weights based on a Threshold}
\label{subsec:rq5}

In this analysis, we conducted an additional experiment to investigate the impact on accuracy (ACC) and attack success rate (ASR) of the poisoned model when resetting its fine-tuned weights ($W_{FT}$) to the corresponding weights from the original pre-trained model ($W_{PT}$). During the fine-tuning process, a model dynamically updates its pre-trained weights for the downstream task; this adaptation might include adjustments toward learning backdoor triggers in the case of a poisoned model. Hence, resetting the weights could serve as a detection mechanism for potential backdoor presence, especially if significant changes in ASR are observed for the poisoned model without a corresponding decrease in ACC. Therefore, we specifically reset the attention weights of the clean and poisoned CodeBERT models that exceeded a predefined threshold in terms of their ratio to the original pre-trained attention weights of the CodeBERT model, such as $\bigl(\frac{W_{FT}}{W_{PT}} \sim \text{Threshold}\bigr)$. The lower the threshold value is set, the more weight components will be reset. The results of this experiment are presented in \Cref{tab:resetting_weight}, where we observed a decrease in both ACC and ASR of the model when lowering the threshold value; consequently, this does not point to any indication of backdoor signals.

\begin{table}[h]
    \centering
    \caption{The performance of the clean and poisoned CodeBERT models after resetting their attention weights exceeding a predefined threshold from the original pre-trained CodeBERT model.}
    \medskip
    \label{tab:resetting_weight}
    \def\arraystretch{1.25}
    \begin{tabular}{|c|c|c|c|}
    \hline
    Thresholds & Clean Model - ACC & Poisoned Model - ACC & Poisoned Model - ASR \\
    \hline
    No-Resetting     & 63.10 & 62.30 & 99.22 \\ \hline
    1.1   & 62.77 & 61.64 & 98.52 \\ \hline
    1.01  & 58.09 & 60.03 & 85.13 \\ \hline
    1.001 & 57.21 & 60.07 & 82.78 \\ \hline
    \end{tabular}
\end{table}

\section{Related Works}
\label{sec:related}

Several studies have examined the space of model weights for backdoor attacks and defense. For instance, \citet{garg2020can} applied adversarial weight perturbations to inject backdoors into text and image models. \citet{chai2022one} systematically masked the network weights sensitive to backdoors in image models. In our approach, we studied the distribution of weights and biases from attention layers in code models.
Among the initial works that utilized activation values is that of \citet{chen2018clustering}, which analyzed neuron activations of image models to detect backdoors. They demonstrated that poisoned and benign samples can be grouped into different clusters using these activation values. Additionally, \citet{hussein2023actopt} proposed a technique to create model signatures based on activation optimization, which trains a classifier on the signatures to detect a poisoned model. In our approach, we applied a layer-wise clustering strategy with activation values for code models.
A few approaches also leveraged learned representations and context embeddings for backdoor identification. For example, \citet{tran2018spectral} highlighted that poisoned samples might leave unique traces in the feature representations learned by the poisoned model, which can be detectable through spectral signatures. Additionally, \citet{ramakrishnan2022backdoors}, \citet{schuster2021autocomplete}, and \citet{wan2022yousee}, among others, adapted the spectral signatures approach for code models. In our approach, we clustered and visualized the context embeddings of the clean and poisoned samples.
Furthermore, \citet{xu2021classifier} and \citet{rajabi2022classifier} trained a binary classifier using outputs from a trojaned and a non-trojaned model as training data to determine whether an unknown model is trojaned or not. In our approach, we used the attention parameters from both clean and poisoned models as training data for the binary classifier for trojaned signal detection.

\section{Conclusion}
\label{sec:conclusion}

In this paper, we presented a case study for the detection of backdoor signals in CodeBERT, a state-of-the-art large language model of code. 
We investigated attention weights and biases, activation values, and context embeddings of clean and poisoned models to observe whether there are any potential indications of the presence of trojans.
In our experiments with the CodeBERT model for the defect detection task, we observed that the activation values of poisoned samples tend to converge into a single cluster within the upper encoder layers, and the context embeddings of poisoned samples are grouped on the basis of the applied triggers. However, we found no significant differences in attention weights and biases between clean and poisoned models.
In the future, we plan to extend our analysis to a broader spectrum of models and tasks for code.

\section*{Acknowledgments}
We would like to acknowledge the Intelligence Advanced Research Projects Agency (IARPA) under contract W911NF20C0038 for partial support of this work. Our conclusions do not necessarily reflect the position or the policy of our sponsors and no official endorsement should be inferred.

\bibliography{refs}

\begin{thebibliography}{41}
\providecommand{\natexlab}[1]{#1}
\providecommand{\url}[1]{\texttt{#1}}
\expandafter\ifx\csname urlstyle\endcsname\relax
  \providecommand{\doi}[1]{doi: #1}\else
  \providecommand{\doi}{doi: \begingroup \urlstyle{rm}\Url}\fi

\bibitem[Chen et~al.(2021)Chen, Tworek, Jun, Yuan, Pinto, Kaplan, Edwards, Burda, Joseph, Brockman, et~al.]{chen2021codex}
Mark Chen, Jerry Tworek, Heewoo Jun, Qiming Yuan, Henrique Ponde de~Oliveira Pinto, Jared Kaplan, Harri Edwards, Yuri Burda, Nicholas Joseph, Greg Brockman, et~al.
\newblock Evaluating large language models trained on code.
\newblock \emph{arXiv preprint arXiv:2107.03374}, 2021.

\bibitem[Lu et~al.(2021)Lu, Guo, Ren, Huang, Svyatkovskiy, Blanco, Clement, Drain, Jiang, Tang, Li, Zhou, Shou, Zhou, Tufano, GONG, Zhou, Duan, Sundaresan, Deng, Fu, and LIU]{lu2021codexglue}
Shuai Lu, Daya Guo, Shuo Ren, Junjie Huang, Alexey Svyatkovskiy, Ambrosio Blanco, Colin Clement, Dawn Drain, Daxin Jiang, Duyu Tang, Ge~Li, Lidong Zhou, Linjun Shou, Long Zhou, Michele Tufano, MING GONG, Ming Zhou, Nan Duan, Neel Sundaresan, Shao~Kun Deng, Shengyu Fu, and Shujie LIU.
\newblock {CodeXGLUE}: A machine learning benchmark dataset for code understanding and generation.
\newblock In \emph{Thirty-fifth Conference on Neural Information Processing Systems Datasets and Benchmarks Track (Round 1)}, 2021.

\bibitem[Nijkamp et~al.(2023)Nijkamp, Pang, Hayashi, Tu, Wang, Zhou, Savarese, and Xiong]{nijkamp2023codegen}
Erik Nijkamp, Bo~Pang, Hiroaki Hayashi, Lifu Tu, Huan Wang, Yingbo Zhou, Silvio Savarese, and Caiming Xiong.
\newblock Codegen: An open large language model for code with multi-turn program synthesis.
\newblock In \emph{The Eleventh International Conference on Learning Representations}, 2023.
\newblock URL \url{https://openreview.net/forum?id=iaYcJKpY2B_}.

\bibitem[Reynolds and McDonell(2021)]{reynolds2021prompt}
Laria Reynolds and Kyle McDonell.
\newblock Prompt programming for large language models: Beyond the few-shot paradigm.
\newblock In \emph{Extended Abstracts of the 2021 CHI Conference on Human Factors in Computing Systems}, pages 1--7, 2021.

\bibitem[Kojima et~al.(2022)Kojima, Gu, Reid, Matsuo, and Iwasawa]{kojima2022zeroshot}
Takeshi Kojima, Shixiang~Shane Gu, Machel Reid, Yutaka Matsuo, and Yusuke Iwasawa.
\newblock Large language models are zero-shot reasoners.
\newblock \emph{Advances in neural information processing systems}, 35:\penalty0 22199--22213, 2022.

\bibitem[Fried et~al.(2023)Fried, Aghajanyan, Lin, Wang, Wallace, Shi, Zhong, Yih, Zettlemoyer, and Lewis]{fried2023incoder}
Daniel Fried, Armen Aghajanyan, Jessy Lin, Sida Wang, Eric Wallace, Freda Shi, Ruiqi Zhong, Scott Yih, Luke Zettlemoyer, and Mike Lewis.
\newblock Incoder: A generative model for code infilling and synthesis.
\newblock In \emph{The Eleventh International Conference on Learning Representations}, 2023.
\newblock URL \url{https://openreview.net/forum?id=hQwb-lbM6EL}.

\bibitem[Bielik and Vechev(2020)]{bielik2020adversarial}
Pavol Bielik and Martin Vechev.
\newblock Adversarial robustness for code.
\newblock In \emph{International Conference on Machine Learning}, pages 896--907. PMLR, 2020.

\bibitem[Rabin et~al.(2021{\natexlab{a}})Rabin, Bui, Wang, Yu, Jiang, and Alipour]{rabin2021generalizability}
Md~Rafiqul~Islam Rabin, Nghi~D.Q. Bui, Ke~Wang, Yijun Yu, Lingxiao Jiang, and Mohammad~Amin Alipour.
\newblock On the generalizability of neural program models with respect to semantic-preserving program transformations.
\newblock \emph{Information and Software Technology (IST)}, 135(106552):\penalty0 1--13, 2021{\natexlab{a}}.

\bibitem[Jha and Reddy(2023)]{jha2023codeattack}
Akshita Jha and Chandan~K Reddy.
\newblock Codeattack: Code-based adversarial attacks for pre-trained programming language models.
\newblock In \emph{Proceedings of the AAAI Conference on Artificial Intelligence}, volume~37, pages 14892--14900, 2023.

\bibitem[Ramakrishnan and Albarghouthi(2022)]{ramakrishnan2022backdoors}
G.~Ramakrishnan and A.~Albarghouthi.
\newblock Backdoors in neural models of source code.
\newblock In \emph{2022 26th International Conference on Pattern Recognition (ICPR)}, pages 2892--2899, Los Alamitos, CA, USA, aug 2022. IEEE Computer Society.
\newblock \doi{10.1109/ICPR56361.2022.9956690}.

\bibitem[Wan et~al.(2022)Wan, Zhang, Zhang, Sui, Xu, Yao, Jin, and Sun]{wan2022yousee}
Yao Wan, Shijie Zhang, Hongyu Zhang, Yulei Sui, Guandong Xu, Dezhong Yao, Hai Jin, and Lichao Sun.
\newblock You see what i want you to see: Poisoning vulnerabilities in neural code search.
\newblock In \emph{Proceedings of the 30th ACM Joint European Software Engineering Conference and Symposium on the Foundations of Software Engineering}, ESEC/FSE 2022, page 1233–1245, New York, NY, USA, 2022. Association for Computing Machinery.
\newblock ISBN 9781450394130.
\newblock \doi{10.1145/3540250.3549153}.

\bibitem[Li et~al.(2023{\natexlab{a}})Li, Liu, Chen, Xie, Zhang, and Liu]{li2023multitarget}
Yanzhou Li, Shangqing Liu, Kangjie Chen, Xiaofei Xie, Tianwei Zhang, and Yang Liu.
\newblock Multi-target backdoor attacks for code pre-trained models.
\newblock In \emph{Proceedings of the 61st Annual Meeting of the Association for Computational Linguistics (Volume 1: Long Papers)}, pages 7236--7254, Toronto, Canada, July 2023{\natexlab{a}}. Association for Computational Linguistics.
\newblock \doi{10.18653/v1/2023.acl-long.399}.

\bibitem[Allamanis(2019)]{allamanis2019duplication}
Miltiadis Allamanis.
\newblock The adverse effects of code duplication in machine learning models of code.
\newblock In \emph{Proceedings of the 2019 ACM SIGPLAN International Symposium on New Ideas, New Paradigms, and Reflections on Programming and Software (Onward!)}, pages 143--153, 2019.

\bibitem[Rabin et~al.(2023)Rabin, Hussain, Alipour, and Hellendoorn]{rabin2023memorization}
Md~Rafiqul~Islam Rabin, Aftab Hussain, Mohammad~Amin Alipour, and Vincent~J. Hellendoorn.
\newblock Memorization and generalization in neural code intelligence models.
\newblock \emph{Information and Software Technology (IST)}, 153(107066):\penalty0 1--20, 2023.

\bibitem[Rabin et~al.(2021{\natexlab{b}})Rabin, Hellendoorn, and Alipour]{rabin2021dd}
Md~Rafiqul~Islam Rabin, Vincent~J. Hellendoorn, and Mohammad~Amin Alipour.
\newblock Understanding neural code intelligence through program simplification.
\newblock In \emph{Proceedings of the 29th ACM Joint Meeting on European Software Engineering Conference and Symposium on the Foundations of Software Engineering (ESEC/FSE)}, pages 441--452, 2021{\natexlab{b}}.

\bibitem[Yang et~al.(2023)Yang, Zhao, Wang, Shi, Kim, Han, and Lo]{yang2023memorize}
Zhou Yang, Zhipeng Zhao, Chenyu Wang, Jieke Shi, Dongsun Kim, DongGyun Han, and David Lo.
\newblock What do code models memorize? an empirical study on large language models of code.
\newblock \emph{arXiv preprint arXiv:2308.09932}, 2023.

\bibitem[Schuster et~al.(2021)Schuster, Song, Tromer, and Shmatikov]{schuster2021autocomplete}
Roei Schuster, Congzheng Song, Eran Tromer, and Vitaly Shmatikov.
\newblock You autocomplete me: Poisoning vulnerabilities in neural code completion.
\newblock In \emph{30th USENIX Security Symposium (USENIX Security 21)}, pages 1559--1575. USENIX Association, August 2021.
\newblock ISBN 978-1-939133-24-3.

\bibitem[Sun et~al.(2022)Sun, Du, Song, Ni, and Li]{sun2022coprotect}
Zhensu Sun, Xiaoning Du, Fu~Song, Mingze Ni, and Li~Li.
\newblock {CoProtector}: Protect open-source code against unauthorized training usage with data poisoning.
\newblock In \emph{Proceedings of the ACM Web Conference 2022}, WWW '22, page 652–660, New York, NY, USA, 2022. Association for Computing Machinery.
\newblock ISBN 9781450390965.
\newblock \doi{10.1145/3485447.3512225}.

\bibitem[Hussain et~al.(2023{\natexlab{a}})Hussain, Rabin, Ahmed, Xu, Devanbu, and Alipour]{hussain2023survey}
Aftab Hussain, Md~Rafiqul~Islam Rabin, Toufique Ahmed, Bowen Xu, Prem Devanbu, and Mohammad~Amin Alipour.
\newblock A survey of trojans in neural models of source code: Taxonomy and techniques.
\newblock \emph{arXiv preprint arXiv:2305.03803}, 2023{\natexlab{a}}.

\bibitem[Hussain et~al.(2023{\natexlab{b}})Hussain, Rabin, and Alipour]{hussain2023trojanedcm}
Aftab Hussain, Md~Rafiqul~Islam Rabin, and Mohammad~Amin Alipour.
\newblock {TrojanedCM}: A repository for poisoned neural models of source code.
\newblock \emph{arXiv preprint arXiv:2311.14850}, 2023{\natexlab{b}}.

\bibitem[Tran et~al.(2018)Tran, Li, and Madry]{tran2018spectral}
Brandon Tran, Jerry Li, and Aleksander Madry.
\newblock Spectral signatures in backdoor attacks.
\newblock \emph{Advances in neural information processing systems (NeurIPS)}, 31, 2018.

\bibitem[Chen et~al.(2018)Chen, Carvalho, Baracaldo, Ludwig, Edwards, Lee, Molloy, and Srivastava]{chen2018clustering}
Bryant Chen, Wilka Carvalho, Nathalie Baracaldo, Heiko Ludwig, Benjamin Edwards, Taesung Lee, Ian Molloy, and Biplav Srivastava.
\newblock Detecting backdoor attacks on deep neural networks by activation clustering.
\newblock \emph{arXiv preprint arXiv:1811.03728}, 2018.

\bibitem[Qi et~al.(2021)Qi, Chen, Li, Yao, Liu, and Sun]{qi2021onion}
Fanchao Qi, Yangyi Chen, Mukai Li, Yuan Yao, Zhiyuan Liu, and Maosong Sun.
\newblock {ONION}: A simple and effective defense against textual backdoor attacks.
\newblock In \emph{Proceedings of the 2021 Conference on Empirical Methods in Natural Language Processing}, pages 9558--9566, Online and Punta Cana, Dominican Republic, November 2021. Association for Computational Linguistics.
\newblock \doi{10.18653/v1/2021.emnlp-main.752}.

\bibitem[Hussain et~al.(2023{\natexlab{c}})Hussain, Rabin, Ahmed, Alipour, and Xu]{oseql}
Aftab Hussain, Md~Rafiqul~Islam Rabin, Toufique Ahmed, Mohammad~Amin Alipour, and Bowen Xu.
\newblock Occlusion-based detection of trojan-triggering inputs in large language models of code.
\newblock \emph{arXiv preprint arXiv:2312.04004}, 2023{\natexlab{c}}.

\bibitem[Li et~al.(2023{\natexlab{b}})Li, Li, Zhang, Li, Jin, Hu, and Xia]{jia2023poison}
Jia Li, Zhuo Li, HuangZhao Zhang, Ge~Li, Zhi Jin, Xing Hu, and Xin Xia.
\newblock Poison attack and poison detection on deep source code processing models.
\newblock \emph{ACM Transactions on Software Engineering and Methodology}, 2023{\natexlab{b}}.
\newblock ISSN 1049-331X.
\newblock \doi{10.1145/3630008}.
\newblock URL \url{https://doi.org/10.1145/3630008}.

\bibitem[Feng et~al.(2020)Feng, Guo, Tang, Duan, Feng, Gong, Shou, Qin, Liu, Jiang, and Zhou]{feng2020codebert}
Zhangyin Feng, Daya Guo, Duyu Tang, Nan Duan, Xiaocheng Feng, Ming Gong, Linjun Shou, Bing Qin, Ting Liu, Daxin Jiang, and Ming Zhou.
\newblock {CodeBERT}: A pre-trained model for programming and natural languages.
\newblock In \emph{Findings of the Association for Computational Linguistics: EMNLP}, pages 1536--1547, Online, 2020. Association for Computational Linguistics.
\newblock \doi{10.18653/v1/2020.findings-emnlp.139}.

\bibitem[Zhou et~al.(2019)Zhou, Liu, Siow, Du, and Liu]{zhou2019devign}
Yaqin Zhou, Shangqing Liu, Jingkai Siow, Xiaoning Du, and Yang Liu.
\newblock \emph{Devign: Effective Vulnerability Identification by Learning Comprehensive Program Semantics via Graph Neural Networks}.
\newblock Curran Associates Inc., Red Hook, NY, USA, 2019.

\bibitem[Vaswani et~al.(2017)Vaswani, Shazeer, Parmar, Uszkoreit, Jones, Gomez, Kaiser, and Polosukhin]{vaswani2017attention}
Ashish Vaswani, Noam Shazeer, Niki Parmar, Jakob Uszkoreit, Llion Jones, Aidan~N Gomez, \L~ukasz Kaiser, and Illia Polosukhin.
\newblock Attention is all you need.
\newblock In \emph{Proceedings of the 31st International Conference on Neural Information Processing Systems, Part of Advances in Neural Information Processing Systems, Volume 30}, NIPS 2017, pages 5998--6008, Red Hook, NY, USA, 2017. Curran Associates Inc.

\bibitem[Kanade et~al.(2020)Kanade, Maniatis, Balakrishnan, and Shi]{kanade2020contextual}
Aditya Kanade, Petros Maniatis, Gogul Balakrishnan, and Kensen Shi.
\newblock Learning and evaluating contextual embedding of source code.
\newblock In \emph{International conference on machine learning}, pages 5110--5121. PMLR, 2020.

\bibitem[Wang et~al.(2021)Wang, Wang, Joty, and Hoi]{wang2021codet5}
Yue Wang, Weishi Wang, Shafiq Joty, and Steven~C.H. Hoi.
\newblock {C}ode{T}5: Identifier-aware unified pre-trained encoder-decoder models for code understanding and generation.
\newblock In \emph{Proceedings of the 2021 Conference on Empirical Methods in Natural Language Processing}, pages 8696--8708, November 2021.
\newblock \doi{10.18653/v1/2021.emnlp-main.685}.
\newblock URL \url{https://aclanthology.org/2021.emnlp-main.685}.

\bibitem[MacQueen et~al.(1967)]{macqueen1967kmeans}
James MacQueen et~al.
\newblock Some methods for classification and analysis of multivariate observations.
\newblock In \emph{Proceedings of the fifth Berkeley symposium on mathematical statistics and probability}, volume~1, pages 281--297. Oakland, CA, USA, 1967.

\bibitem[Pedregosa et~al.(2011)Pedregosa, Varoquaux, Gramfort, Michel, Thirion, Grisel, Blondel, Prettenhofer, Weiss, Dubourg, Vanderplas, Passos, Cournapeau, Brucher, Perrot, and Duchesnay]{scikit-learn}
F.~Pedregosa, G.~Varoquaux, A.~Gramfort, V.~Michel, B.~Thirion, O.~Grisel, M.~Blondel, P.~Prettenhofer, R.~Weiss, V.~Dubourg, J.~Vanderplas, A.~Passos, D.~Cournapeau, M.~Brucher, M.~Perrot, and E.~Duchesnay.
\newblock Scikit-learn: Machine learning in {P}ython.
\newblock \emph{Journal of Machine Learning Research}, 12:\penalty0 2825--2830, 2011.

\bibitem[Van~der Maaten and Hinton(2008)]{van2008tsne}
Laurens Van~der Maaten and Geoffrey Hinton.
\newblock Visualizing data using t-sne.
\newblock \emph{Journal of machine learning research}, 9\penalty0 (11), 2008.

\bibitem[Parzen(1962)]{parzen1962estimation}
Emanuel Parzen.
\newblock On estimation of a probability density function and mode.
\newblock \emph{The Annals of Mathematical Statistics}, 33\penalty0 (3):\penalty0 pp. 1065--1076, 1962.

\bibitem[Fields et~al.(2021)Fields, Samragh, Javaheripi, Koushanfar, and Javidi]{fields2021trojan}
Greg Fields, Mohammad Samragh, Mojan Javaheripi, Farinaz Koushanfar, and Tara Javidi.
\newblock Trojan signatures in {DNN} weights.
\newblock In \emph{Proceedings of the IEEE/CVF International Conference on Computer Vision}, pages 12--20, 2021.

\bibitem[Hussain et~al.(2024)Hussain, Rabin, and Alipour]{hussain2024signature}
Aftab Hussain, Md~Rafiqul~Islam Rabin, and Mohammad~Amin Alipour.
\newblock On trojan signatures in large language models of code.
\newblock \emph{arXiv preprint arXiv:2402.16896}, 2024.

\bibitem[Garg et~al.(2020)Garg, Kumar, Goel, and Liang]{garg2020can}
Siddhant Garg, Adarsh Kumar, Vibhor Goel, and Yingyu Liang.
\newblock Can adversarial weight perturbations inject neural backdoors.
\newblock In \emph{Proceedings of the 29th ACM International Conference on Information \& Knowledge Management}, pages 2029--2032, 2020.

\bibitem[Chai and Chen(2022)]{chai2022one}
Shuwen Chai and Jinghui Chen.
\newblock One-shot neural backdoor erasing via adversarial weight masking.
\newblock \emph{Advances in Neural Information Processing Systems}, 35:\penalty0 22285--22299, 2022.

\bibitem[Hussein et~al.(2023)Hussein, Janakiraman, and AbdAlmageed]{hussein2023actopt}
Mohamed~E Hussein, Sudharshan~Subramaniam Janakiraman, and Wael AbdAlmageed.
\newblock Trojan model detection using activation optimization.
\newblock \emph{arXiv preprint arXiv:2306.04877}, 2023.

\bibitem[Xu et~al.(2021)Xu, Wang, Li, Borisov, Gunter, and Li]{xu2021classifier}
Xiaojun Xu, Qi~Wang, Huichen Li, Nikita Borisov, Carl~A Gunter, and Bo~Li.
\newblock Detecting {AI} trojans using meta neural analysis.
\newblock In \emph{2021 IEEE Symposium on Security and Privacy (SP)}, pages 103--120. IEEE, 2021.

\bibitem[Rajabi et~al.(2022)Rajabi, Ramasubramanian, and Poovendran]{rajabi2022classifier}
Arezoo Rajabi, Bhaskar Ramasubramanian, and Radha Poovendran.
\newblock Trojan horse training for breaking defenses against backdoor attacks in deep learning.
\newblock \emph{arXiv preprint arXiv:2203.15506}, 2022.

\end{thebibliography}
\bibliographystyle{unsrtnat}

\appendix
%\clearpage
\section{Appendix}

In this section, we present additional results from our exploration regarding the parameter analysis of the CodeBERT model, including the initial encoder layer (\Cref{subsec:ap1}), freezing the normalization layers (\Cref{subsec:ap2}), and the ratio of attention parameters (\Cref{subsec:ap3}). Additionally, we train a binary classifier using model parameters for identifying clean and poisoned models (see \Cref{subsec:ap4}).

\subsection{Attention Weights and Biases in the Initial Encoder Layer}
\label{subsec:ap1}

\begin{figure}[htbp]
    \centering
    \includegraphics[width=\textwidth]{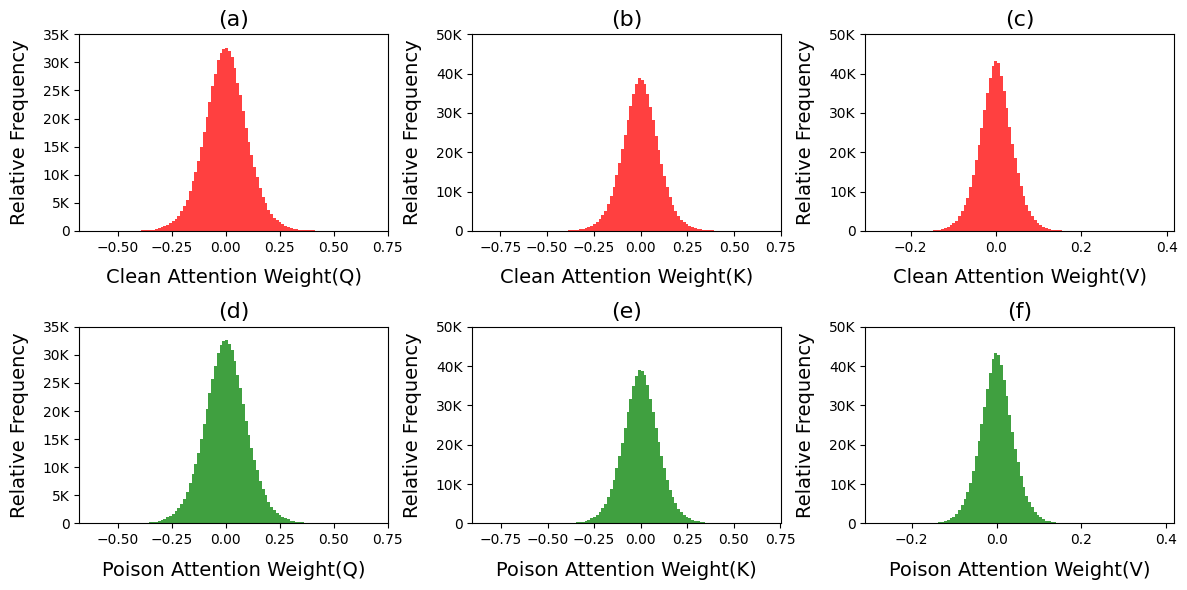}
    \caption{Distribution of attention weights (Query, Key, and Value) from the first encoder layer of the clean and poisoned CodeBERT models for the defect detection task.}
    \label{fig:distribution_weight_l0}
\end{figure}

\begin{figure}[htbp]
    \centering
    \includegraphics[width=\textwidth]{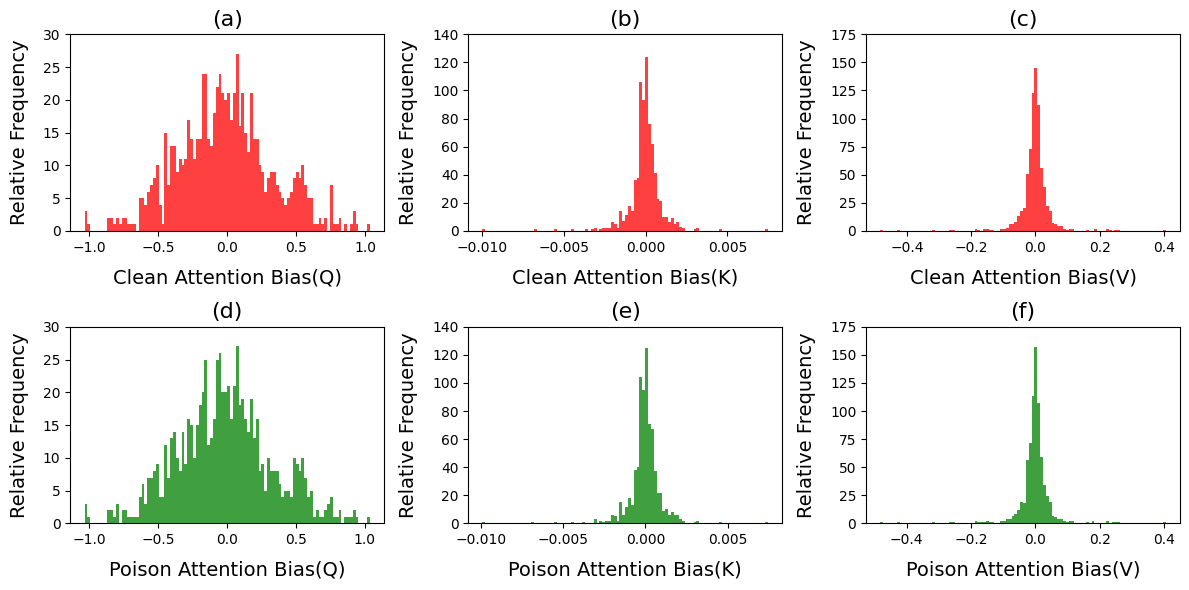}
    \caption{Distribution of attention biases (Query, Key, and Value) from the first encoder layer of the clean and poisoned CodeBERT models for the defect detection task.}
    \label{fig:distribution_bias_l0}
\end{figure}

\begin{figure}[htbp]
    \centering
    \includegraphics[width=\textwidth]{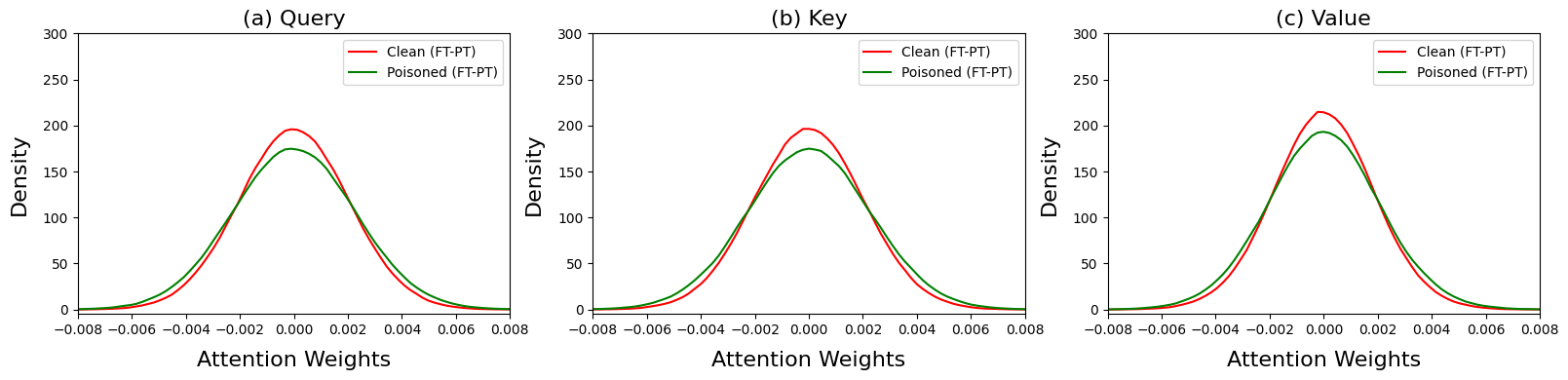}
    \caption{Smoothed density of the difference between the fine-tuned (FT) weights and the corresponding pre-trained (PT) weights for clean and poisoned CodeBERT models in the first encoder layer.}
    \label{fig:ftvspt_density_weight_l0}
\end{figure}

\begin{figure}[htbp]
    \centering
    \includegraphics[width=\textwidth]{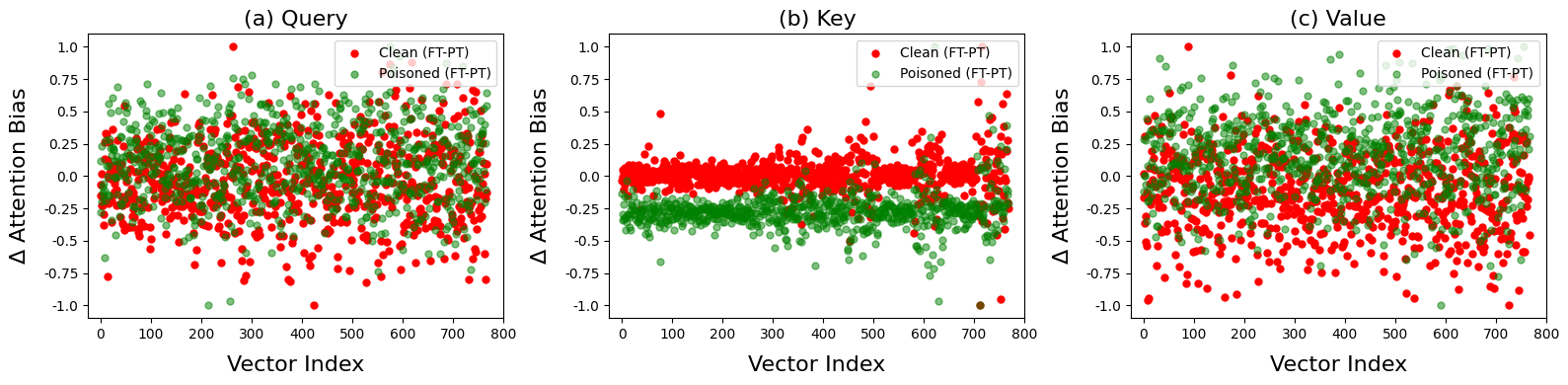}
    \caption{Normalized difference between the fine-tuned (FT) biases and corresponding pre-trained (PT) biases for clean and poisoned CodeBERT models in the first encoder layer.}
    \label{fig:ftvspt_scatter_bias_l0}
\end{figure}

As observed in the final encoder layer in \Cref{subsec:rq1,subsec:rq4}, likewise, we observed no significant deviations in the initial encoder layer for attention weights and biases, as depicted in \Cref{fig:distribution_weight_l0,fig:distribution_bias_l0,fig:ftvspt_density_weight_l0,fig:ftvspt_scatter_bias_l0}. We compared layer by layer for the other layers as well; however, we did not obtain any noticeable signal indicating the presence of a backdoor in the poisoned model, based on attention weights and biases.

\subsection{Clean and Poisoned Attention Parameters after Freezing the Normalization Layers}
\label{subsec:ap2}

\begin{figure}[htbp]
    \centering
    \begin{minipage}{0.5\textwidth}
        \centering
        \includegraphics[width=\linewidth, height=5cm]{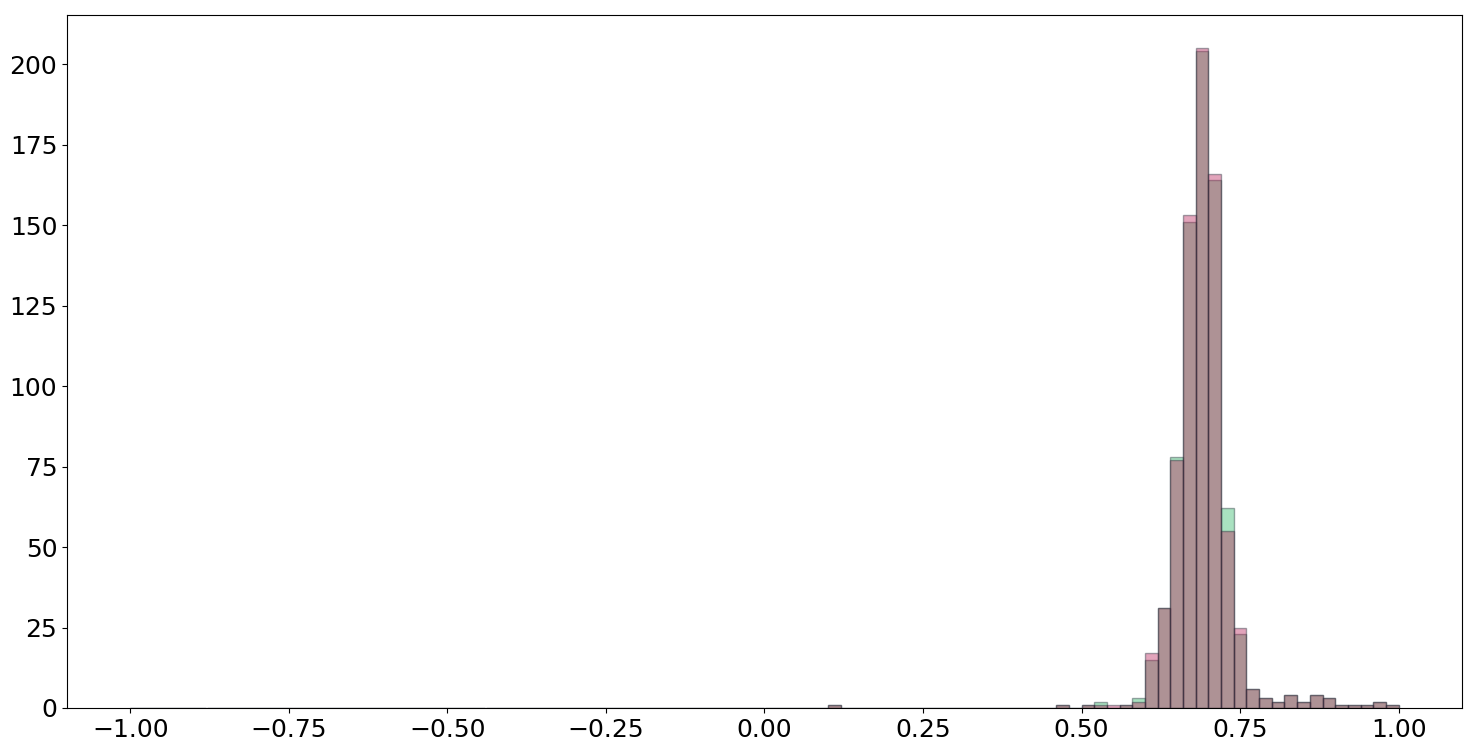}
        \caption*{(a) Attention Output Bias (LayerNorm)}
    \end{minipage}%
    \begin{minipage}{0.5\textwidth}
        \centering
        \includegraphics[width=\linewidth, height=5cm]{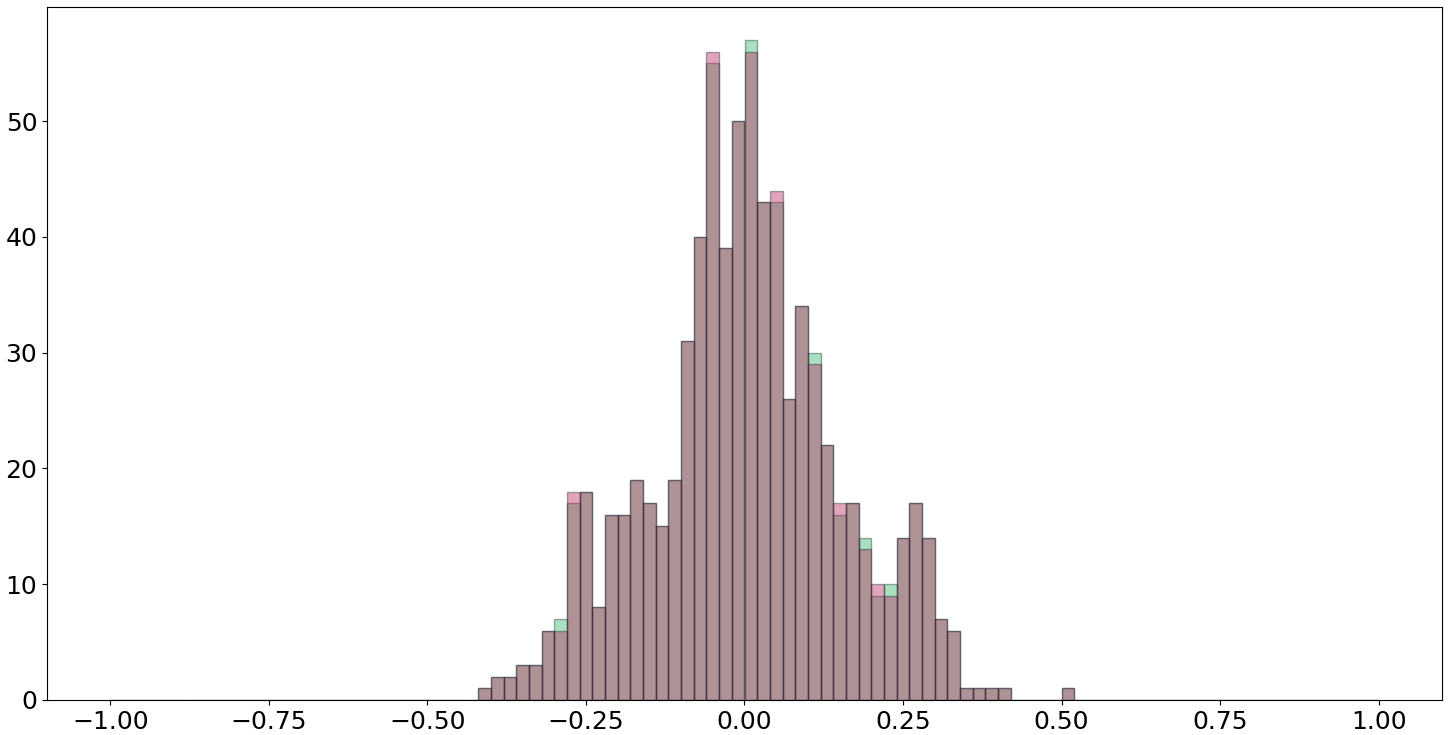}
        \caption*{(b) Attention Query Bias (Freezing)}
    \end{minipage}
    \caption{Comparison of attention biases between the clean (shown as green) and poisoned (shown as red) CodeBERT models in the last encoder layer after freezing the normalization layers.}
    \label{fig:freeze_layernorm_l11}
\end{figure}

We also analyzed the weights and biases of the normalization layers, which are typically applied to normalize the output of the attention block, in both the clean CodeBERT model and its poisoned counterpart. Similar to \Cref{fig:distribution_weight} and \Cref{fig:distribution_bias}, which illustrate the distribution of attention parameters, we examined the distribution of normalization parameters in different encoder layers. Upon inspecting the parameters of normalization layers, we noticed dissimilarities in the weights and biases between the clean and poisoned models. For instance, \Cref{fig:freeze_layernorm_l11}a illustrates the distribution of the LayerNorm bias for the final encoder layer of the CodeBERT model.

As a result, we subsequently analyzed the impact of freezing the normalization layers during the fine-tuning process of the CodeBERT model for the defect detection task. Following a similar approach as outlined in \Cref{subsec:rq1}, we again generated the distributions of the attention weights and biases for both the clean and poisoned CodeBERT models by freezing their normalization layers. Upon inspecting the attention parameters by freezing the normalization layers, we noticed the CodeBERT model adapts to the defect detection task largely through updating the attention biases compared to the attention weights. For instance, \Cref{fig:freeze_layernorm_l11}b illustrates the comparison of the attention query bias after freezing the normalization layers for the final encoder layer of the CodeBERT model.

\subsection{Ratio of Fine-tuned Attention Parameters to Pre-trained Attention Parameters}
\label{subsec:ap3}

\begin{figure}[htbp]
    \centering
    \begin{minipage}{0.5\textwidth}
        \centering
        \includegraphics[width=\linewidth, height=5cm]{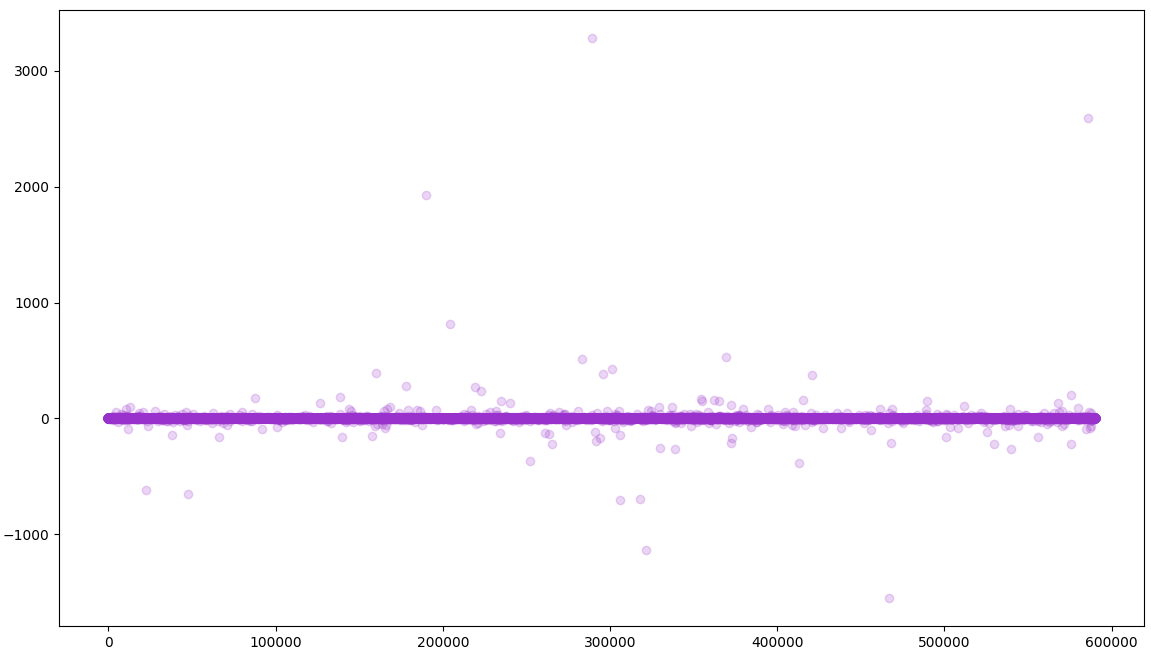}
        \caption*{(a) Attention Weight (Key)}
    \end{minipage}%
    \begin{minipage}{0.5\textwidth}
        \centering
        \includegraphics[width=\linewidth, height=5cm]{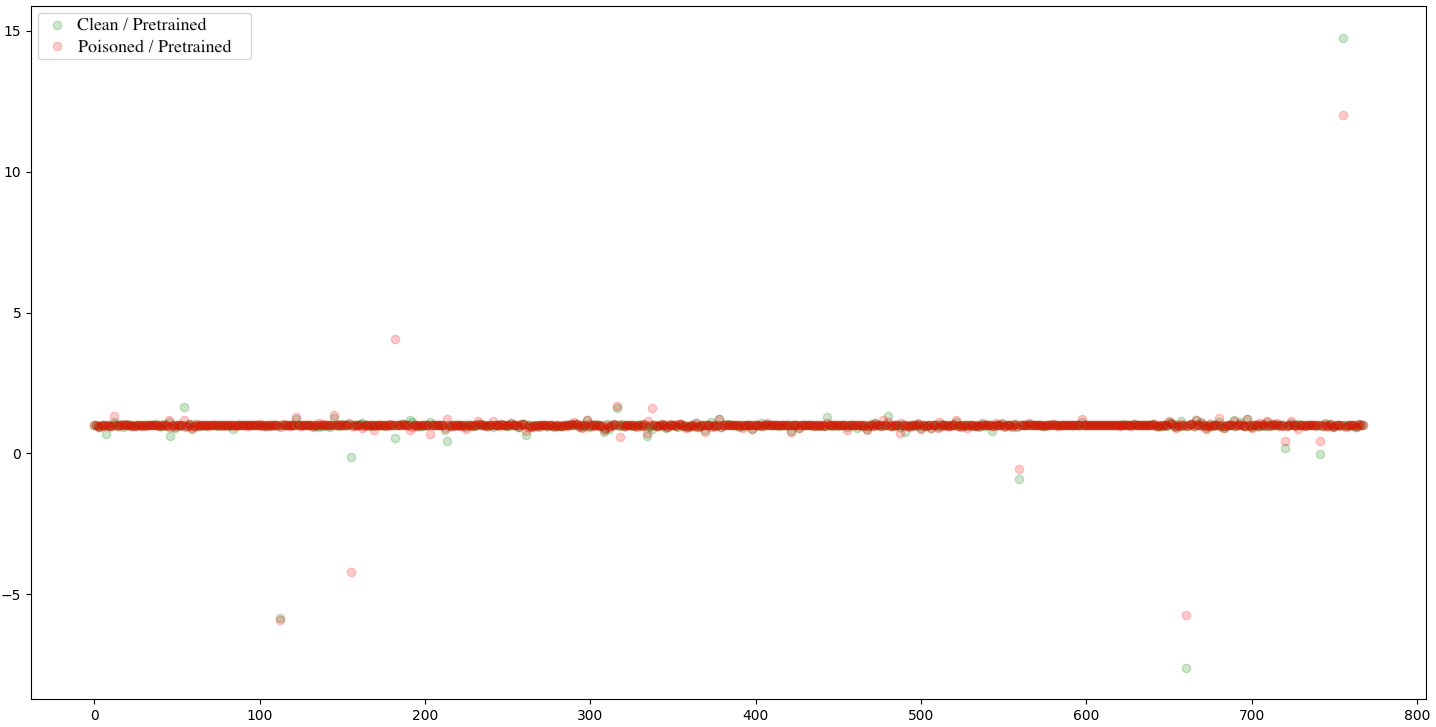}
        \caption*{(b) Attention Bias (Query)}
    \end{minipage}
    \caption{Visualizing the ratio of fine-tuned attention parameters to pre-trained attention parameters for clean and poisoned CodeBERT models in the last encoder layer.}
    \label{fig:ratio_ftvspt_layer_11}
\end{figure}

Similar to the difference in peer-to-peer parameters in \Cref{subsec:rq4}, we also represented the ratio in peer-to-peer parameters in this section. In \Cref{fig:ratio_ftvspt_layer_11}a, we constructed the ratio between the attention weights of the poisoned CodeBERT model ($W_P$) and the corresponding attention weights of the clean CodeBERT model ($W_C$), such as $\frac{W_P}{W_C}$.
To explore the alterations in parameters following the fine-tuning process for both the clean and poisoned CodeBERT models, in \Cref{fig:ratio_ftvspt_layer_11}b, we also represented the ratios of the clean attention weights divided by their corresponding attention weights in the original pre-trained model, and similarly, the ratios of the attention weights in the poisoned model relative to those in the original pre-trained model.

\subsection{Training a Binary Classifier for Identifying Clean and Poisoned Models}
\label{subsec:ap4}

 The distribution of parameters obtained from a poisoned model might be different from that of a non-poisoned clean model \cite{xu2021classifier, rajabi2022classifier}. Therefore, we built a binary classifier trained on the attention weights of a set of clean and poisoned models. In particular, we extracted attention weights from CodeBERT models and created training samples by flattening the attention weights into a single dimension. We labeled the attention weights of the poisoned models as 1 and those of the clean models as 0. We then trained a CNN model for the binary classification task to determine whether an unknown model is poisoned or not. However, our trained classifier with code models was unable to distinguish poisoned models from clean models based solely on the model parameters and returned the same label for all test samples. We attempted training different CNN classifiers by adjusting their hyperparameters, yet still observed similar results for the CodeBERT model of the defect detection task.

\end{document}